&pdflatex

\documentclass[aip,pop,amsmath,amssymb,amsfonts, reprint,preprintnumbers]{revtex4-1}

\usepackage{graphicx}
\usepackage{dcolumn}
\usepackage{bm}
\usepackage[mathlines,columnwise]{lineno}

\usepackage[utf8]{inputenc}
\usepackage[T1]{fontenc}
\usepackage{mathptmx}
\usepackage{tabularx}
\usepackage{xcolor}

\usepackage{tcolorbox}

\usepackage[normalem]{ulem}

\usepackage{catchfile}
\usepackage{xstring}

\DeclareUnicodeCharacter{FFFD}{XXXX Here I am?????XXXX}

\begin{document}

	\title{The data-driven future of high energy density physics}

	\author{Peter W. Hatfield}
	\email{peter.hatfield@physics.ox.ac.uk}
	\affiliation{Clarendon Laboratory, University of Oxford, Parks Road, Oxford OX1 3PU, UK}

	\author{Jim A. Gaffney}
	\email{gaffney3@llnl.gov}
	\affiliation{Lawrence Livermore National Laboratory, 7000 East Ave, Livermore, CA 94550, USA}

	\author{Gemma J. Anderson}
	\email{anderson276@llnl.gov}
	\affiliation{Lawrence Livermore National Laboratory, 7000 East Ave, Livermore, CA 94550, USA}

	\author{Suzanne Ali}
	\affiliation{Lawrence Livermore National Laboratory, 7000 East Ave, Livermore, CA 94550, USA}

	\author{Luca Antonelli}
	\affiliation{York Plasma Institute, Department of Physics, University of York, Heslington, York YO10 5DD, UK}

	\author{Suzan Ba\c{s}e\u{g}mez du Pree}
	\affiliation{Nikhef, National Institute for Subatomic Physics, Science Park 105, 1098 XG, Amsterdam, Netherlands}

	\author{Jonathan Citrin}
	\affiliation{DIFFER - Dutch Institute for Fundamental Energy Research, Eindhoven, Netherlands}

	\author{Marta Fajardo}
	\affiliation{Instituto de Plasmas e Fus\~ao Nuclear, Instituto Superior T\'ecnico, 1049-001 Lisboa, Portugal}

	\author{Patrick Knapp}
	\affiliation{Sandia National Laboratories, 1515 Eubank SE, Albuquerque, New Mexico 87185-1196, USA}

	\author{Brendan Kettle}
	\affiliation{Imperial College London, London, SW7 2AZ, UK}

	\author{Bogdan Kustowski}
	\affiliation{Lawrence Livermore National Laboratory, 7000 East Ave, Livermore, CA 94550, USA}

	\author{Michael J. MacDonald}
	\affiliation{Lawrence Livermore National Laboratory, 7000 East Ave, Livermore, CA 94550, USA}

	\author{Derek Mariscal}
	\affiliation{Lawrence Livermore National Laboratory, 7000 East Ave, Livermore, CA 94550, USA}

	\author{Madison E. Martin}
	\affiliation{Lawrence Livermore National Laboratory, 7000 East Ave, Livermore, CA 94550, USA}

	\author{Taisuke Nagayama}
	\affiliation{Sandia National Laboratories, 1515 Eubank SE, Albuquerque, New Mexico 87185-1196, USA}

	\author{Charlotte A.J. Palmer}
	\affiliation{School of Mathematics and Physics, Queen's University Belfast, Belfast, BT7 1NN, UK}

	\author{J. L. Peterson}
	\affiliation{Lawrence Livermore National Laboratory, 7000 East Ave, Livermore, CA 94550, USA}

	\author{Steven Rose}
	\affiliation{Clarendon Laboratory, University of Oxford, Parks Road, Oxford OX1 3PU, UK}
	\affiliation{Imperial College London, London, SW7 2AZ, UK}

	\author{JJ Ruby}
	\affiliation{Laboratory for Laser Energetics, University of Rochester, Rochester, NY, USA}

	\author{Carl Shneider}
	\affiliation{Dutch National Center for Mathematics and Computer Science (CWI), Science Park 123, 1098 XG, Amsterdam, Netherlands}

	\author{Matt J.V. Streeter}
	\affiliation{Imperial College London, London, SW7 2AZ, UK}

	\author{Will Trickey}
	\affiliation{York Plasma Institute, Department of Physics, University of York, Heslington, York YO10 5DD, UK}

	\author{Ben Williams}
	\affiliation{AWE Plc, Aldermaston, Reading, RG4 7PR, United Kingdom}


	\date{\today}


	\pacs{}
	\keywords{}


	\begin{abstract}
	\end{abstract}


  \maketitle


\section*{Preface}

\textbf{The study of plasma physics under conditions of extreme temperatures, densities and electromagnetic field strengths is significant for our understanding of astrophysics, nuclear fusion and fundamental physics. These extreme physical systems are strongly non-linear and very difficult to understand theoretically or optimize experimentally. Here, we argue that machine learning models and data-driven methods are in the process of reshaping our exploration of these extreme systems that have hitherto proven far too non-linear for human researchers. From a fundamental perspective, our understanding can be helped by the way in which machine learning models can rapidly discover complex interactions in large data sets. From a practical point of view, the newest generation of extreme physics facilities can perform experiments multiple times a second (as opposed to $\sim$daily) – moving away from human-based control towards automatic control based on real-time interpretation of diagnostic data and updates of the physics model. To make the most of these emerging opportunities, we advance proposals for the community in terms of research design, training, best practices, and support for synthetic diagnostics and data analysis.  }

\section*{ }

`Set the controls for the heart of the Sun' encouraged a 2004 paper\cite{Rose2004} (riffing on the 1968 Pink Floyd song), describing the bright future of using Earth based experiments to create conditions similar to inside the Sun in the lab. Seventeen years later substantial advances have been made in this research programme. A question that is presently emerging however is \textit{who} should be at the controls - humans, or artificial intelligences?

In the last few years plasma physics has been steadily beginning to explore the use of modern day data science and artificial intelligence methods to support research goals\cite{Spears2018,Wang2020}. In this article we will identify data science issues for physics specifically at the extreme\cite{Colvin2013}; extremely high temperatures, densities or electromagnetic field strengths - which have unique challenges. In particular phenomena at these conditions are highly \textit{non-linear} - small parameter changes can lead to large changes in behaviour. Interpreting extreme physics data typically requires simultaneously comprehending large amounts of complex multi-modal data from multiple different sources. Optimising extreme physics systems requires fine-tuning over large numbers of (often highly correlated) parameters. Artificial Intelligence (AI) methods have proved highly successful at teasing out correlations in large data sets like these and we believe will be crucial for understanding and optimising systems that up to now have been inscrutable.   These extreme conditions can be found in astrophysical scenarios, but can also be created using high-energy `drivers' (often lasers) in the laboratory - millimetre sized plasmas with temperatures and pressures higher than the centre of the Sun. The field has seen an explosion of interest in machine learning techniques because new and future laser facilities have much higher shot (and corresponding data) rates than previous facilities. Data from laboratory experiments can help us understand astrophysical plasmas, let us probe new phenomena in particle physics that conventional accelerators can't reach, and might even lead the way to nuclear fusion as a power source. In this paper we will highlight what data science issues are relevant for extreme plasma science, discuss successes in the field, identify what challenges remain, and look towards the future.

One of the most challenging areas of extreme plasma physics is High Energy Density Physics (HEDP), a sub-field dating back to the 1940s that seeks to understand the behaviour of macroscopic matter that is simultaneously at \textit{both} very high temperatures and pressures; typically >$10^7$ K and >$10^6$ bar. At these conditions, several complex areas of physics become relevant and highly coupled, making \textit{ab initio} predictions very challenging. For example, in many circumstances HEDP plasmas can start to inherit properties from both `classical plasmas' (where the behaviour of the matter can to some degree be thought of as a gas of both ions and free electrons) and condensed matter (matter at solid density where strong interactions between bound electrons are relevant)\cite{Graziani2014}. Key contemporary problems in HEDP theory include understanding multi-species plasmas, self-consistent emission, absorption, and scattering of radiation, non-equilibrium plasmas, relativistic electron transport, magnetised plasmas, and quantum electrodynamic effects.

Understanding HEDP is of both great theoretical and practical importance. As already discussed, understanding these conditions is key in astrophysics. The field is also a ripe domain for new areas of fundamental physics; it is becoming possible to study particle and nuclear physics through HEDP experiments, as well as novel phenomena that are predicted to only emerge at extreme conditions e.g. the predicted thermal Schwinger process - spontaneous production of electrons and positrons at very high electric field strengths\cite{Gould2019}. Extreme physics and high power laser science have given access to exotic forms of matter e.g. new forms of ice\cite{Millot2018} and metallic hydrogen\cite{Celliers2018}. HEDP experiments are also at the forefront of the development of new classes of particle accelerators\cite{Joshi2010} (e.g. Laser Wakefield acceleration\cite{Hidding2019}, bright gamma ray sources\cite{Wang2018}, laser-driven ion acceleration\cite{Badziak2018} and highly efficient neutron generation\cite{Feng2020}) with wide ranging application across science including condensed matter physics, material science, and biomedical imaging\cite{Wakefield2019}. Finally, the high temperatures and pressures of HEDP are one way to make nuclear fusion as a clean industrial power source a reality via \textit{inertial confinement fusion}\cite{nuckolls72} (ICF).

HEDP has a rich heritage of experimentation, currently practiced by thousands of scientists in several large facilities around the world. The National Ignition Facility (NIF) at Lawrence Livermore National Laboratory (LLNL) is the most energetic laser in the world, and is the premier facility working towards ICF\cite{Hurricane2016}, as well as operating an exciting Discovery Science programme\cite{Fournier2019,Remington2015}. At the other end of the energy spectrum are high repetition rate lasers (for example Gemini at the Central Laser Facility, CLF\cite{CLF2019}), which are much less energetic (and so can typically reach less extreme conditions), but can fire up to many times a second rather than at most a few times a day at NIF. There are many more facilities, each with unique capabilities, and a range of other technologies that are used around the world in HEDP experiments, including gas guns, Z-pinches, proton beams, pulsed power, X-Ray Free-Electron Lasers (XFELs) and ion accelerators (e.g. the Facility for Antiproton and Ion Research\cite{Sturm2011}). New facilities and upgrades are constantly in planning, and understanding the data we currently have is directly relevant to choices about what facilities we will need in the future. There are also important synergies with closely related areas of physics, for example magnetic confinement fusion (MCF), solar probes, and the detection of high energy cosmic rays. Finally, alongside experimentation, huge computational facilities have traditionally been a key part of HEDP, with the development of many sophisticated simulation codes run on top high performance computing (HPC) facilities. At the time of writing, $\sim6$ of the top ten supercomputers in the world are used in some capacity for simulating plasma physics experiments like ICF.

\begin{figure}[!h]
\centering\includegraphics[width=3in]{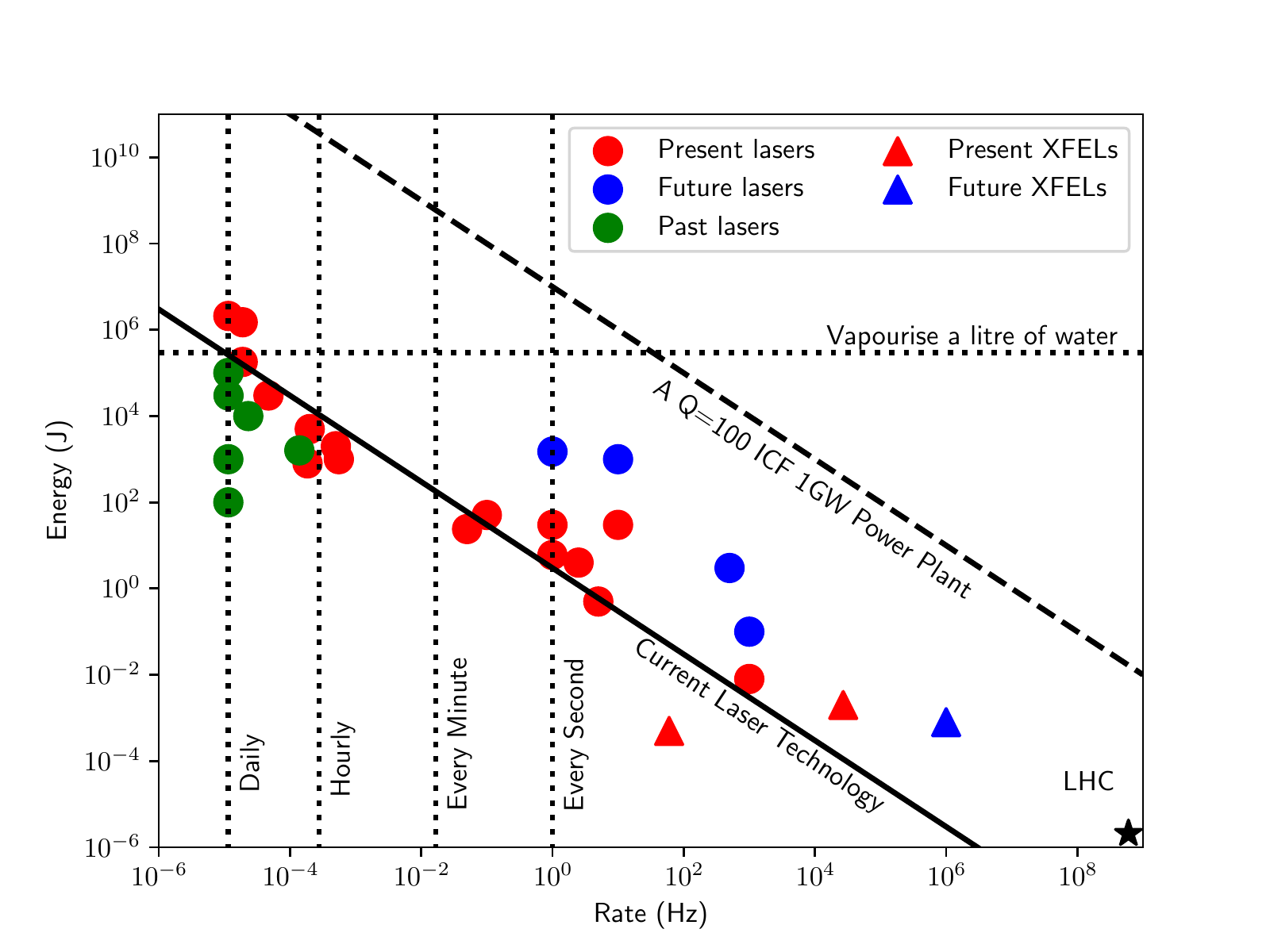}

\caption{Shot rates and energy of large high powered laser facilities in different eras. Shot rates and energies plotted are representative rather than definitive; facilities typically can operate in a number of slightly different modes, and in addition other laser properties (long pulse versus short pulse, beam colour etc.) are also important. Note also that higher shot numbers enable more parameter space to be probed, or a higher signal-to-noise ratio to be reached, but do not automatically translate into more data (depending on what diagnostics are used and the experiment performed). Facilities included are: NIF, LMJ, Omega, Gemini, Vulcan, Orion, TA2, Artemis, Vega, Titan, Texas Petawatt, HAPLS, SG-III, CoReLS 4PW, RRCAT 150TW, LULI2000, ALLS, Shanghai Superintense Ultrafast Laser Facility, DRACO (current), Allegra, EPAC DiPOLE, TARANIS-X, Station of Extreme Light (future), Nova, SHIVA, Cyclops, Argus, SG-II, ISKRA-V (past)\cite{Roso2018,Zheng2016,Danson2019,LaserOpportunities2018,Lin1999,LULI2019,Kirillov1990,Zhao:LINAC2018-MO2A01,Zhang2020,Schramm2017}, SACLA, the European XFEL (current XFELs) and LCLS-II (a future XFEL)\cite{EXFEL,LCLSII,Yabashi2017}. The solid line illustrates the approximate state of current technology. We show for context the collision rate, and energy per collision for the LHC\cite{CERN2017}, although such figures are not directly comparable. We also indicate with a dashed line the shot rates that would be need to be achieved for an ICF power plant\footnote{For $Q=100$; an output energy to input energy ratio of 100 per shot}, to illustrate the long term aspirations of the field.}
\label{fig_shot_rates}
\end{figure}

The field is entering the regime where it is necessary to systematically manage large quantities of data, both because the amount of experimental data is set to massively grow, and also because the capacity to simulate huge numbers of experiments is moving beyond the limits of conventional methods. The quantity of experimental data is increasing both because shot rates on facilities are dramatically increasing (see Figure \ref{fig_shot_rates}), but also because diagnostics are becoming more sophisticated; $\sim$150GB of data is taken on each NIF shot, and LCLS campaigns have reached $\sim$70GB per minute\cite{MacDonald2016}. Machine learning\cite{Mitchell1997}, Bayesian methods\cite{Sivia2006}, and data-driven science\cite{Brunton2019} have been common in particle physics and astrophysics for many years, and are having large impacts on other multiscale, highly non-linear areas of physics e.g. climate science and Earth system science\cite{Reichstein2019}\cite{Fleming2020}. Some AI solutions from other fields are likely to be applicable in plasma physics - but HEDP also has its own unique challenges. Specifically we typically want to (sometimes very rapidly) fine tune (either optimising or fitting a model) a large number of parameters for a desired outcome, based on a large number of multi-modal data sets. This is very difficult for humans, as it is very hard to simultaneously comprehend all the different sources and forms of data - but is achievable for AI. In HEDP, the highest performing experiments in the field are now increasingly data driven\cite{Streeter2018,Gopalaswamy2019}; the vision is to work towards a HEDP science where algorithms are at the centre of design, analysis of experiments, and discovery.

In the following sections we review key challenges and topics in extreme plasma physics, and highlight research areas where data science has dramatically impacted the field. We further lay out what the future might hold for data driven Extreme Physics and HEDP - and how the community must change and adapt how it practices research to make the most of these exciting new approaches.


\section*{Challenges}

\begin{table*}[t!] \label{tab:methods}
  \footnotesize
{\renewcommand{\arraystretch}{1.6}
\setlength{\tabcolsep}{7pt}
  \begin{tabularx}{\textwidth}{>{\hsize=.7\hsize}X X X X X}
   \hline
  Analytical Task & Scientific Task & Conventional approaches & Limitations of conventional approaches & Emergent or potential approaches \\
    \hline
    \hline
       \textbf{Uncertainty \newline Quantification}
     & Quantifying uncertainty on estimate of some microphysics (e.g. opacity, equation-of-state), characterising uncertainty estimate on laser energy required for ignition & Simulation-based sensitivity studies, basic Poisson/Gaussian uncertainties, not including correlations & Local sensitivities only, \newline relies on estimated uncertainties in underlying parameters, \newline cannot account for simulation bias & MCMC, dropout, bootstrapping, quantification of `unknown unknowns', invertible neural networks \\
     \hline
       \textbf{Regression}
    & Emulating an expensive simulation, building an empirical model of microphysics & Look-up tables, polynomial regression & Struggles in high dimension, incorporating known physics constraints, extrapolation is difficult, often not fast enough & Neural networks, Gaussian processes, autoencoders \\
    \hline
    \textbf{Design}
    & Selecting target design parameters, scheduling observation runs  & Adjusting parameters by hand, using a combination of code output and Designer judgement & Hard in high dimensions, human intensive, cannot be done quickly, can miss optimal/novel designs & Bayesian optimisation, Genetic/Evolutionary algorithms \\
    \hline
    \textbf{Pattern \newline Recognition}
    & Identifying target defects, characterising magnetic perturbations, image segmentation/featurization & Human inspection & Laborious and time
consuming, subject to individual biases, uncontrolled approximation to true information content & Convolutional Neural Networks, Deep learning, random forest classifiers, Human-AI hybrid approaches, Application Specific Integrated Circuits incorporated into diagnostics, Generative adversarial networks \\
    \hline
    \textbf{Data \newline Synthesis}
    & Combining data from multiple sources (e.g. different instruments) & Researcher guided inference from independently analysed diagnostic data & Difficult to do `by hand', hard to take advantages of any degeneracies broken  & Bayesian inference, data assimilation methods \\
    \hline
    \textbf{Classification}
    & Image classification, particle track classification, identify good/bad shots, anomaly detection & Simple analytic criteria, human inspection & Laborious and time consuming, potentially inaccurate  & Random forest, decision trees, neural networks, deep learning, Generative adversarial networks \\
    \hline
    \textbf{Model Calibration}
    & Update physics models/parameters in the face of experimental data & Trial and error, single-point fitting to data, hand-tuning & Laborious and time consuming, inaccurate or missing treatment of uncertainties, miss multiple solutions, prone to overfitting & Bayesian inference, discrepancy modeling, transfer learning, multi-task learning, physics informed neural networks\\
  \end{tabularx}}
\caption{Conventional and AI-enabled approaches to tasks in extreme plasma physics}
\end{table*}

The qualities that make HEDP an exciting area of research also contribute to causing the challenges inherent in any quantitative analysis of data. Table I summarises some key quantitive methods, comparing conventional and emerging approaches, and in the following sub-sections we discuss three primary challenges that data science techniques can help address in HEDP.

\subsection*{Experimental Design and Automation}

A key component of HEDP is experimentation. Designing experiments is a hugely complex task, and researchers typically will have a wide range of overlapping goals during this phase. Researchers must consider what specific hypotheses are to be tested, and if the expected data would be sufficient to rule out alternatives. What design to shoot, diagnostic instruments to field, or astronomical observation to make will depend heavily on the specific science goal at hand. In current design approaches there is typically much intuition implicitly at play, and often experiments are built as an extension of what has been done before, limiting the regions of experimental space that are studied. Machine learning techniques offer a possible framework for the intuition of the computational scientists and experimental scientists to be explicitly included in a cohesive picture that considers both measurements that can be made, and which aspects of physics have the most leverage on those measurements. AI-aided design is beginning to be used in the creation of new HEDP experiments \cite{Hatfield2019, Martin2018}, and we foresee that becoming the norm in the coming years. However, machine learning methods have yet to demonstrate that they can ``think outside the box'' of pre-defined parameter spaces, therefore for the foreseeable future it will still be necessary to have substantial human input in the design process.

Experiments on state-of-the-art high repetition rate lasers, firing many times a second, cannot be done with a human in the loop - so in this case some algorithmic control is essential. This in principle could lead to huge savings of time, money, and human effort in the near future - allowing them to be redirected to aspects of research where they can be better used. Automating experiments combines control of experimental parameters and real-time analysis of experimental results into a single algorithm. The experimental goals are coded in to the automation, such that choices of how to vary the experimental inputs are made automatically in order to maximise the output.

Automation also allows for active feedback stabilisation of complex processes. This is of particular benefit to HEDP experiments, where many non-linear effects combine to determine the performance of what is a nominally unpredictable process. In this context, AI could be the solution to an otherwise intractable problem; the well-known ability of deep learning models to discover complex interactions\cite{Raghu2020} could be used to optimize systems that have proven far too non-linear for human researchers.

\subsection*{Data Synthesis}

The measurements made in HEDP experiments are often highly integrated; experiments typically don't measure the actual quantity of interest (QoI), and there are usually multiple confounding or nuisance variables that have to be controlled. Isolating a particular aspect of these systems is often not possible, resulting in a measurement of an evolving system subject to different conditions and physical processes. This complexity makes repeatability an issue. In order to isolate individual aspects of the underlying physics, experiments therefore typically require multiple, indirect observations, sometimes spread across several different experimental facilities. At most facilities researchers have developed multiple diagnostics for experiments; for example both x-ray and particle spectra may be measured on a single experiment, along with many other forms of experimental data - all of which might contribute to the determination of a single quantity. The analysis of such increasingly sophisticated interlinked data requires the use of more advanced modelling techniques to make the best use of the available data, and to sensibly quantify the uncertainty on any inferences. Looking forward, large quantities of data will require the development of streamlined automated data analysis tools to avoid read/write bottlenecks\cite{Thayer2017}.

As well as combining data from multiple diagnostics, there are also challenges in combining data from multiple sources; `data synthesis', where multiple forms of heterogeneous data are combined in a self-consistent manner in order to construct a more complete picture of the phenomenon of interest. Typically each diagnostic will be analysed separately, and overall conclusions reached heuristically by the researcher. However, combining all available data can help reduce error bars, break degeneracies, and cancel out uncorrelated systematics\cite{Bernal2018}.

The long term vision for the best use of the physics data is to develop systems to combine data from multiple diagnostics on the same shot, multiple shots, shots on different facilities, and finally from different types of facility.

\subsection*{Physics models}

The evolution of HEDP experiments is governed by multiple complex, non-linear physics models, each of which has its own range of applicability and uncertainties. Solving these multiphysics computer models requires very expensive simulations, which are often not suited to next-generation HPC platforms. With the increasing desire to explore larger experimental parameter spaces, there is a shifting dynamic between single, incredibly computationally expensive, `hero' simulations, and large-scale ensembles of simulations which only become meaningful when confronted with experimental data.

Due to numerical approximations, poorly known or unknown model parameters, and missing physics (\emph{model discrepancy}), computer models often do not accurately represent the physical process under study. We can leverage real-world experiments to calibrate our computational models, enabling us to constrain some of the uncertain model parameters; ideally including an Uncertainty Quantification (UQ) analysis\cite{Osthus2019} and practicing `data assimilation'\cite{Lang2019} that obeys physical laws. A useful approach to this problem is through \emph{Bayesian inversion} (also known as model calibration)~\cite{Gaffney2019,Kasim2019}, which allows prior knowledge to be be included and for which convenient numerical tools exist. In a HEDP context, challenges with this approach include large parameter spaces, expensive models, and very sparse experimental data.

The computer models can often be prohibitively computationally expensive.
In this case a {\it surrogate model (or emulator)} may be useful - running a moderate number of expensive simulations, and training a machine learning algorithm to reproduce what the simulation would have given as an output. Surrogate models are of course themselves only an approximation of the true model, introducing further uncertainty that needs to be accounted for.

Emulation can be done at the macro level (e.g. predicting outputs of a whole experiment), or at the level of individual modules run inline inside of a computer model. The use of emulators opens up an array of new inference methods that would not be practical with the full computational expense of a conventional simulation\cite{Plassche2020, Meneghini2017, Anirudh2020,Kluth2020,Humbird2018, Kasim2020,Kustowski2020}.


\section*{Case Studies}

Here we highlight three key areas where researchers are tackling the challenges described in Section II, and where data science is significantly impacting the practice of extreme plasma science.

\subsection*{Astrophysics}

Plasmas are found throughout the Universe: in Solar physics (the centre of the Sun, the solar corona, solar wind); interplanetary, interstellar and intergalactic media; in Earth's and other planetary magnetospheres and ionospheres, and tails of comets; in compact astrophysical objects (white dwarfs, neutron stars, and black holes) and their accretion disks. After direct matter-antimatter annihilation, accretion onto compact objects is the most efficient energy source in the Universe\cite{Bambi2018}; the Cosmos offers plenty of opportunity to probe extreme physics.

Understanding our closest star is of course of supreme practical importance - alongside curiosity-driven blue sky astrophysical motivation. Space weather refers to conditions on the Sun, in the heliosphere, in the solar wind, and in Earth's magnetosphere, ionosphere, and thermosphere, that can influence the performance and reliability of space-borne and ground-based technological systems and can endanger human life or health\cite{SpaceWeather1995,Eastwood2017}. Changes in the space environment, resulting mainly from changes on the Sun, include modification of the ambient plasma, particulate radiation (electrons, protons and ions), electromagnetic radiation (including radio, visible, UV, X-ray, and gamma radiation), and magnetic and electric fields.

As with many other areas of physics, the amount of data we have on the multi-scale complex physics experiment that is the Sun has massively increased in recent years from Solar space missions e.g. from both spacecraft, such as the Solar and Heliospheric Observatory (SOHO), the Hinode mission, the Solar Dynamics Observatory (SDO), the Parker Solar Probe (PSP) and Solar Orbiter (SolO), and CubeSats, such as the Miniature X-ray Solar Spectrometer (MinXSS). Machine learning has been employed successfully to both forecast and `now-cast' space weather\cite{Camporeale2019a}. It has been used to make predictions and gain insight about: solar wind\cite{Camporeale2017}, solar flares\cite{Chen2019,Campi2019}, coronal mass ejections (CMEs)\cite{Inceoglu2018,Bobra2016}, Van Allen radiation belts\cite{Sarma2020}, geomagnetically induced currents\cite{Camporeale2020} and the role of auroras as proxy for ionospheric disturbances\cite{Lamb2019}.

\begin{figure*}
\includegraphics[width=7in]{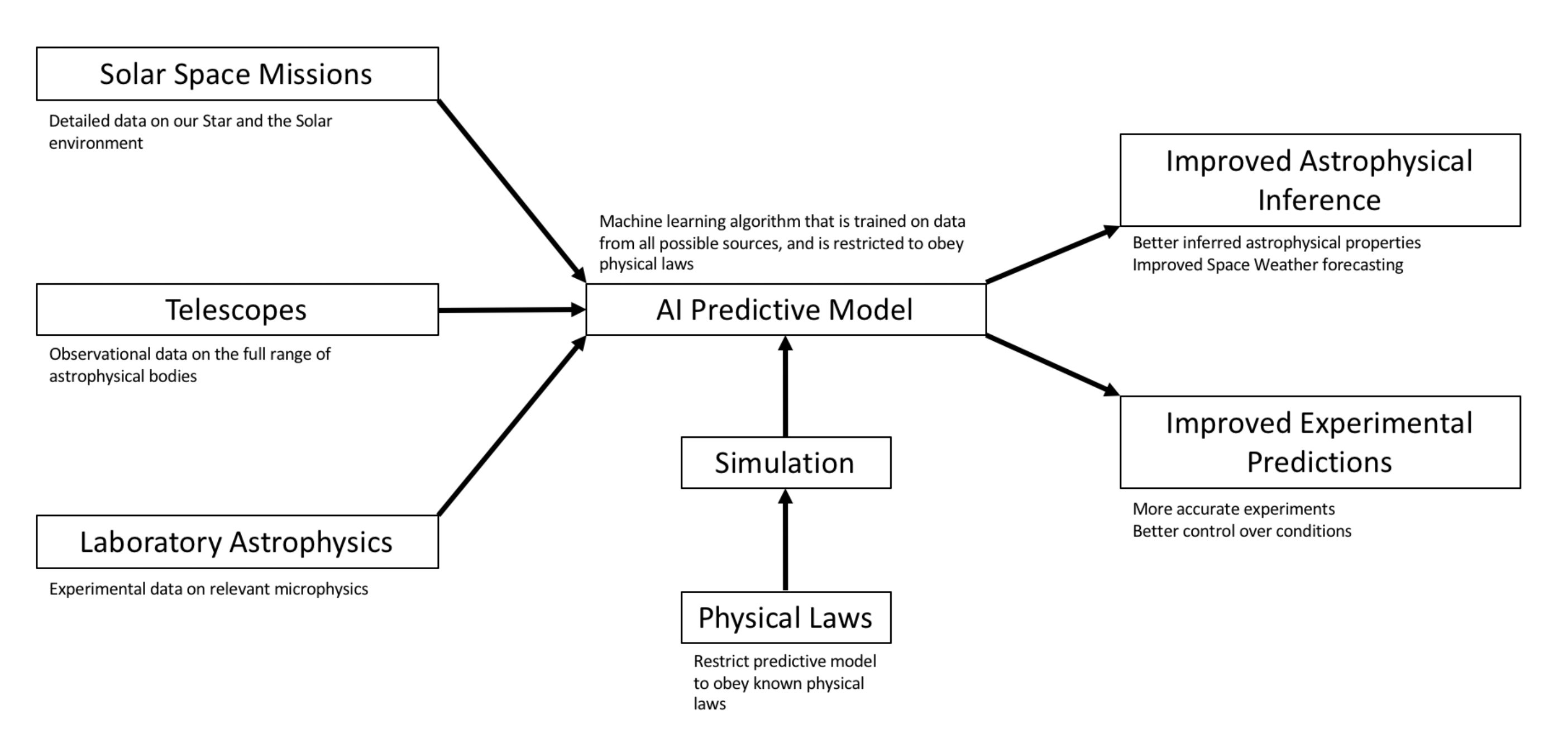}
\caption{\textbf{Integration of astrophysical information:} combining data from multiple sources. At the centre is a machine learning algorithm that is receiving data from multiple sources, and is able to update its beliefs based on this (`data assimilation'). The resulting data-driven results will take information from both theory, and data, to give more precise predictions, with realistic uncertainties.}
\label{fig_astro}
\end{figure*}

Outside our Solar System, large survey telescopes are taking huge amounts of data on astronomical bodies with extreme physics. A full range of data science based techniques are used to both a) identify objects of interest in the large data sets\cite{Rowlinson2019}, and b) understand their underlying physics. Machine learning and statistical methods have been used to: make a Bayesian constraint on supra-nuclear equations-of-state\cite{Coughlin2019}, understand the interiors of exoplanets\cite{Dorn2017}, constrain fundamental stellar parameters from asteroseismic observations\cite{Bellinger2016}, classify supernovae\cite{Lochner2016}, identify and infer the properties of white dwarfs\cite{Kong2018}, classify states of black hole X-ray binaries\cite{Huppenkothen2017}, and emulate radiative transfer during the epoch of cosmological reionization\cite{Chardin2019}.

HEDP experimental facilities can, as discussed, also probe the extreme plasma physics relevant in astrophysical bodies. For example, experiments have measured the equation of state at conditions relevant for the centres of gas giant exoplanets\cite{Saumon2004,Smith2014}, tested theories on possible origins of magnetic fields on galactic scales\cite{Tzeferacos2018} and helped our understanding of white dwarf photosphere spectra \cite{Falcon2015}. There has recently been great success in applying data science methods to these experiments and making inferences with realistic uncertainties. Measurements of iron opacity at conditions relevant to solar physics on the Z facility for example have modelled a large number of possible sources of uncertainty and systematics, and have combined the data from multiple shots together to get realistic uncertainties. This measurement was statistically inconsistent with conventional opacity predictions and has potentially led to an adjustment of estimates of the metallicity (lithium and higher atomic number elements) of the Sun\cite{Bailey2015,Nagayama2019}. Similarly, Bayesian estimates from experiment of nuclear reaction rate at conditions relevant for Big Bang nucleosynthesis were found to be 3\% different from conventionally assumed; a level of precision needed for cosmological studies\cite{DeSouza2019}. As discussed in the introduction, future facilities will have shot rates that will make multi-shot studies like this the norm. Critically this will make it possible to consistently make realistic uncertainty quantifications of key parameters, and will also give the ability to probe large parts of parameter space (e.g. measure the equation-of-state or opacity at a large number of points in temperature-density space). With these vast quantities of observational and experimental data, the natural next step is to use experimentally calibrated models of microphysics like these in astrophysical models - with the potential to give improved predictions over theory-only models.

There is a huge amount of data on astrophysical plasmas of a wide variety of sources, taken in a huge variety of ways, but unfortunately they are currently held in different forms, by different communities. See Figure \ref{fig_astro} for an infogram on how different astrophysical data sets could in future be combined in a data assimilation framework\cite{Zhelavskaya2017,Zhelavskaya2018}. Astronomers have long practiced \textit{`multi-wavelength' astronomy} (astronomy observing at multiple electromagnetic wavelengths), and since $\sim$2015 have practiced \textit{`multi-messenger' astronomy} (astronomy taking data from multiple messengers of information, electromagnetic waves, gravitational waves, cosmic rays and neutrinos). It is now time for \textit{`multi-provenance' astronomy}; integrating data from both observations and experiments into one coherent model of our understanding of astrophysical sources.

\subsection*{Inertial Confinement Fusion}

The aim of igniting a self-propagating hydrogen fusion reaction in laboratory-scale plasmas has been a scientific grand challenge for over 50 years. The motivation is clear: a reliable fusion reaction could form the basis for an effectively limitless, clean, and safe energy source \cite{freidberg2008plasma}. There are good reasons the research has been long-running; generating densities and temperatures high enough to overcome the Coulomb repulsion between nuclei and Bremsstrahlung radiation losses from electrons, over timescales long enough to allow energy break-even, is extremely difficult. In nature, the required conditions are reached in the cores of stars as a result of gravitational collapse and confinement. On Earth, we require even more extreme conditions to account for the much shorter time scales.\par

In inertial confinement fusion (ICF) studies, millimetre-scale targets are driven to implode using high energy (100s kJ to MJ-class) drivers \cite{atzeni2004physics}. For fusion ignition, the implosion and subsequent stagnation needs to generate a high temperature ($10^7$-$10^8$ K) hydrogen plasma, surrounded by a compressed ($100\times$ - $1000\times$) fuel layer, confined by its own inward velocity for several hundred picoseconds \cite{lindl1998inertial}. Experiments on current ICF facilities like the National Ignition Facility (NIF) \cite{campbell99,Moses_2008}, the Omega Laser Facility \cite{BOEHLY1997495}, and the Z pulsed-power machine \cite{Slutz:2012aa,gomez2014} routinely generate plasmas under solar core conditions and beyond. \par

ICF experiments present distinct data challenges due to their scale and complexity. Experimental facilities are expensive and are not expected to achieve high repetition-rate operation any time soon. Targets and drivers are very complex resulting in high-dimensional experimental design spaces. Experiments are also highly \emph{integrated}, meaning direct measurement of any figure of merit beyond the raw energy yield is not possible. These factors mean that ICF datasets are always sparse, with multiple confounding factors and uncertain information content; researchers therefore place a very high value on theoretical studies undertaken using multiphysics simulation codes. While cheaper than experiments, the simulations are still expensive, requiring at least CPU-months to complete. They can also have significant bias\cite{Dimonte2020}, and therefore require calibration against the available experimental data (from both ICF, and smaller scale experiments focused on relevant phenomena). There is a significant need for new methods that can help with experimental design and optimization, interpretation of experimental data, linking experiments with physics models, as well as making reliable predictions of future experiments. As this work progresses, ICF is becoming a prototypical example of the difficulties associated with science in the data-poor regime.\par

\begin{figure*}
 \includegraphics[width=7in]{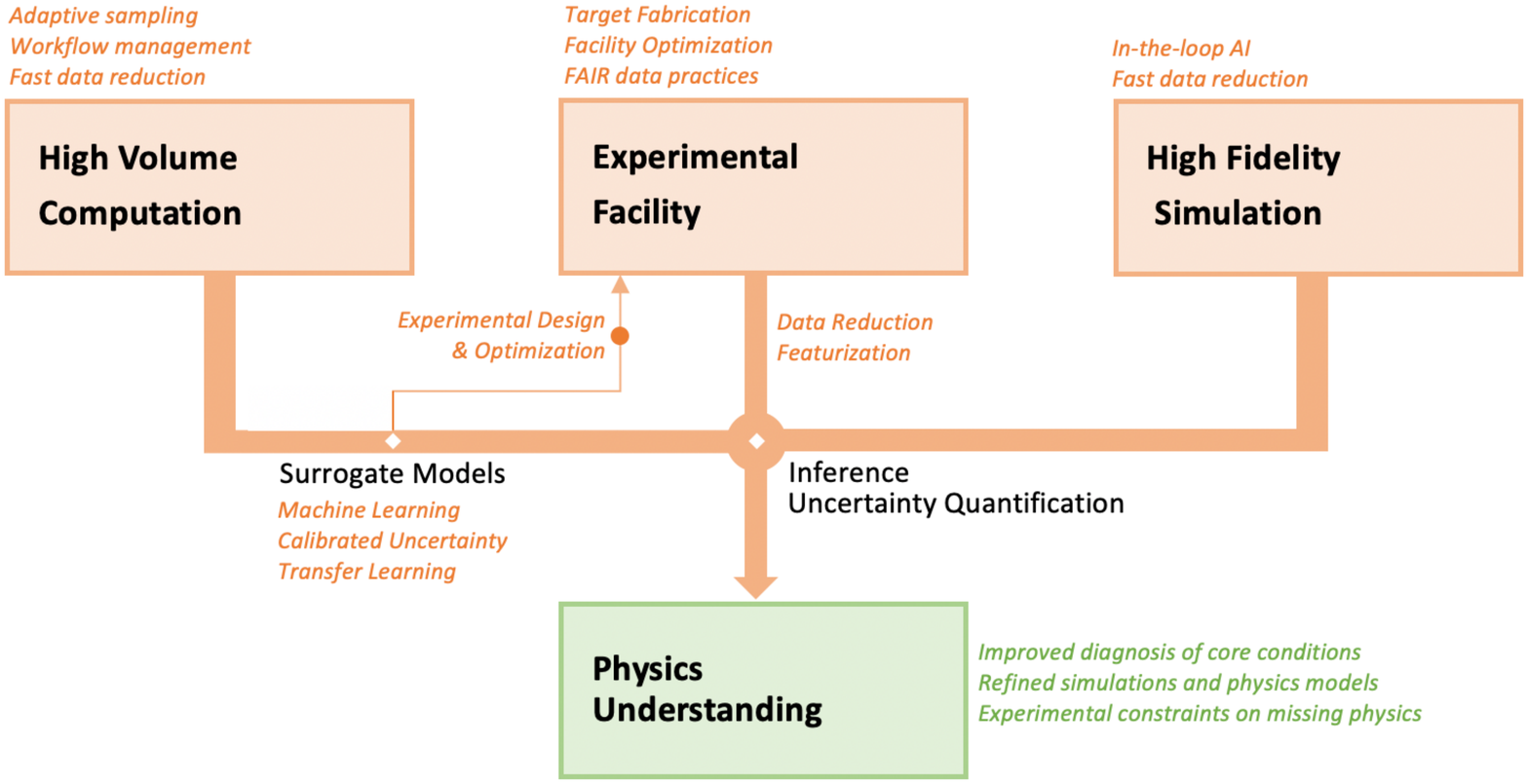}
 \caption{\textbf{Integrating Information Sources in ICF Studies:} Our understanding of ICF implosion physics is based on a combination of high-volume, lower-fidelity simulation ensembles; sparse, difficult-to-diagnose experiments; and best-physics simulations that push the limits of high-performance computing technology. Creating and synthesizing these data into an improved understanding of the physics will require multiple new, complementary techniques from data science, uncertainty quantification (UQ) and artificial intelligence.
 }
 \label{fig_icf}
 \end{figure*}

 As with the other examples in this perspective, the fundamental data problem is the synthesis of multiple sources of information. Here, the key aim is to efficiently use the sparse information available to update our physics understanding in order to make simulations more predictive of future experiments. Figure \ref{fig_icf} shows a potential workflow that fully integrates data-driven and machine-learning methods to achieve this goal. There are three fundamental information sources; experimental data which may comprise $\sim10$ `shots' each producing multiple diagnostics with diverse data types; traditional high fidelity simulations which produce single best-physics predictions for each shot; and high-volume ensemble studies which use large numbers of (necessarily lower-fidelity) simulations to investigate competing physics hypothesis and provide protection against overfitting. The optimization and control of these information sources, combination of the resulting data, and updating of models to become more predictive present numerous opportunities for modern data science and machine learning approaches and each stage has seen recent active research.\par

Work to optimize the three information sources in Figure \ref{fig_icf} include the acceleration of multiphysics simulations using deep learning \cite{yang_2019,Kluth2020}, infrastructure for intelligent control of large-scale simulations \cite{peterson2019merlin}, intelligent and data-informed design of experiments \cite{Peterson2017,Hatfield2019,Gopalaswamy2019}, as well as optimization of experimental facilities \cite{Amorin2019}. These developments have enabled simulation studies of unprecedented size \cite{nora17} and the generation of open-source ICF datasets \cite{jag_dataset_2020} that motivate novel deep learning research \cite{Anirudh2020,Thiagarajan2020DesigningAE,Kasim2020,Hatfield2020}. AI tools have been applied to the automatic analysis and featurization of complex data types like spectra, images\cite{Glinsky_2019, Anirudh2020}, and line-of-sight dependent quantities. There has been significant interest in using Bayesian inference to improve diagnostics \cite{Palaniyappan_2020}, and to synthesize observations in both focused HEDP experiments \cite{Kasim2019} and full-scale ICF experiments \cite{gaffney_2013,GAFFNEY2013457,Knapp2018,Gaffney2019}. The ultimate aims of using these methods to improve physics understanding, and the reliability of simulations in extrapolating to new designs\cite{Osthus2019} or facilities, have been addressed though machine learning \cite{Gopalaswamy2019,Hsu_2020}, Bayesian model calibration \cite{Gaffney2019}, and transfer learning \cite{Humbird2020,Kustowski2020}. \par

Integrating the recent work we have described into a fully developed workflow similar to Figure \ref{fig_icf} is still a significant challenge. Once achieved, however, we expect to gain an unprecedented view of the ICF design space, significantly better understanding of the conditions in current ICF experiments, and an improved understanding of the path to high yield and ignition. \par

\subsection*{Automation for High Repetition Rates}

\begin{figure*}
\centering\includegraphics[width=7in]{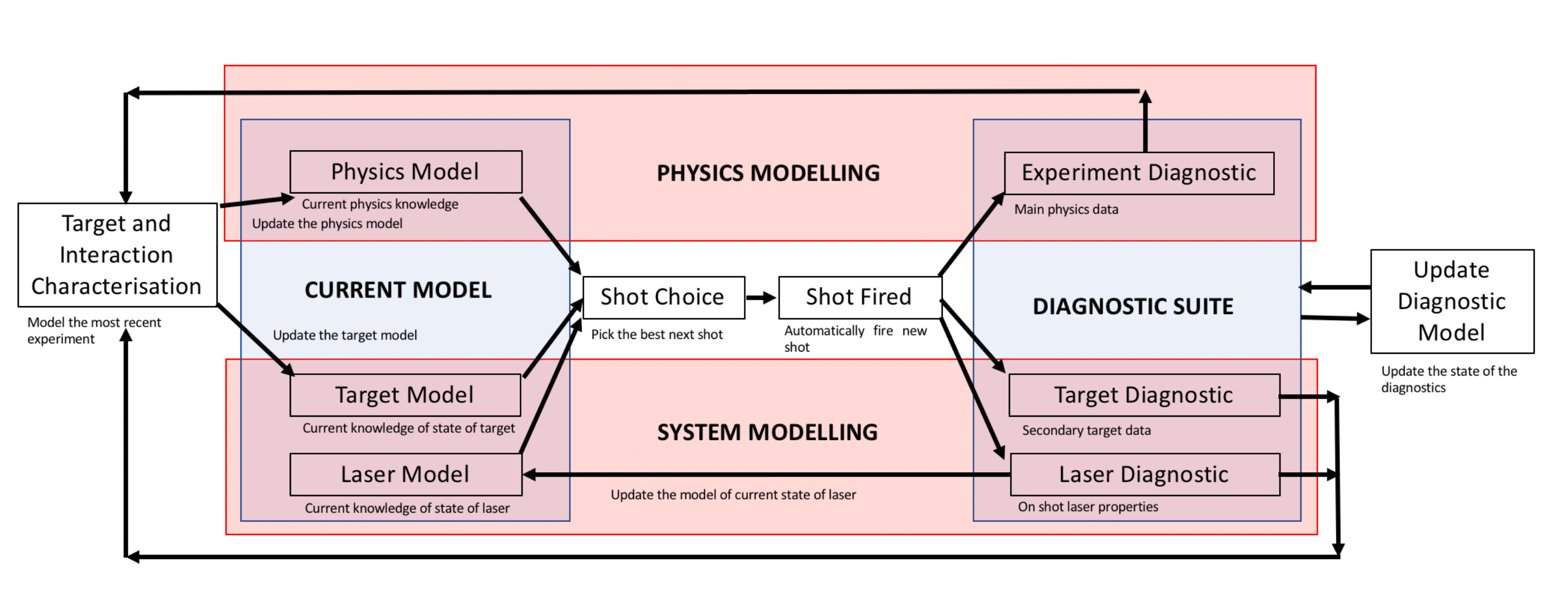}
\caption{\textbf{High repetition workflow:} the different components of a series of high repetition rate high powered laser. The AI system must have i) a model of its best estimate, with uncertainties, of the physics being probed, ii) a model of the current state of the laser (both so it can better achieve science goals, but also to avoid damaging itself), iii) a model of the target, iv) an algorithm to rapidly select what the next shot must be (depending on science goal), v) a system to actually fire the shot with no human intervention, vi) rapid automated data collection, vii) rapid physics, laser and target modelling, and viii) capacity to update a model of diagnostic performance (if needed). }
\label{fig_high_rep}
\end{figure*}

At the other end of the spectrum to NIF, which can only fire roughly once a day, are high-repetition rate lasers that can fire up to multiple times a second. This means huge amounts of data towards given science goals, but also means large aspects of the experimental process must be automated. To succeed, the automation of experiments requires both control of experimental parameters \textit{and} real-time analysis of experimental results in one single algorithmic process; see Figure \ref{fig_high_rep}.

The first step in achieving this is for experimental goals to be coded in to the automation algorithm, such that choices of how to vary the experimental inputs are made automatically in order to maximise the desired output.  This approach enables huge increases in efficiency in optimisation and model learning experiments, especially important when these experiments are resource limited. In addition, automation also allows for active feedback stabilisation of complex processes; this is of particular benefit to high energy density and plasma physics experiments, where many non-linear (and also often non-equilibrium) effects combine to determine the performance of what is a nominally unpredictable process. Thought must also be given to other factors that will determine optimal data return; diagnostics used, signal-to-noise ratio achieved, and experimental parameter space covered. Many diagnostics have already been adapted for fast electronic readout\cite{Gotzfried2018}, however, challenges still remain in adapting x-ray or charged particle imagers and similar detectors for both maximum flexibility, as well as robustness to the harsh radiation and debris environments of HEDP experiments - especially at multi-Hz repetition rates.

This approach may be used to perform the following tasks:
\begin{itemize}
    \item \textit{Optimisation}: A function of experimental diagnostics is used to calculate the `fitness' which expresses how closely measurements reflect the desired performance.  An iterative procedure is then performed to optimise this fitness value by controlling experiment input parameters.  This can be done with any optimisation procedure e.g. an evolutionary approach (i.e. genetic algorithms \cite{He2015,Streeter2018,Dann2019}), a numerical minimisation method (e.g. Nelder-Mead \cite{Dann2019}) or by Bayesian optimisation using a machine learned surrogate model \cite{Kirschner2019}\cite{Shalloo2020}.  This approach can also be used to limit the parameter search to satisfy some safety constraint, e.g. beam loss in a  particle accelerator \cite{Kirschner2019}. In general applying these approaches allows for rapid optimisation of experiments in a far more efficient manner than human controlled experiments - and produces much better results.

    \item \textit{Stabilisation}: Active feedback can improve the stability of experimental performance by rapidly controlling input parameters to counteract oscillation or drifts in the apparatus\cite{Maier2020}. This is routinely performed to stabilise component systems, such as alignment of laser beam transport, but can also be applied to highly complex and non-linear experimental phenomena, such as density limit disruptions in tokamaks using the predictions of neural network \cite{Jtext2018,Fu2020,Kates-Harbeck2019}.  A stable output source then allows for much better experimental or source application outcomes.

    \item \textit{Model Inference}: A Bayesian approach to statistical inference and model validation requires incorporating experiment uncertainties from diagnostic data in a rigorous manner, accounting for correlations across all parameter spaces\cite{Gaffney2019}. This can lead not only to better estimates of the uncertainty, but the results of the inference can also dramatically change. Including this approach in real time in the data taking process can be used to ensure that the experiment optimally constrains the physical models under examination - getting the most \textit{information} per shot (as opposed to simply the most neutrons/x-rays etc.).

\end{itemize}

Encoding the scientific goals into the algorithm making the shot choice is just one of many  challenges in automating high repetition rate facilities. Not only the physics of interest, but also the laser system and the target setup typically have inherent non-linearities that can make automating knowledge extraction extremely challenging. Small changes in system parameters (laser pulse width, shape, energy, focal spot conditions, target thickness etc.) can lead to large changes in experimental outcomes - requiring very fine control of the entire system. Thus the laser itself and the target must be modelled alongside the physics of interest. This complex multi-modal data must also be analysed as fast as the shot-rate to prevent another bottleneck in the experimental loop. In addition performance of diagnostics themselves might be impaired over time (for example if exposed to large radiation fluxes), requiring further modelling. The goal is for the AI to understand the effect of these diagnosed (and potentially undiagnosed) fluctuations in the system, rather than be confused by it. Human intuition risks misinterpreting evidence when many parameters are changing simultaneously. Finally, data archiving from experiments will rapidly become a challenge. Significant challenges in developing pipelines that can prevent data bottlenecks will become important, i.e. the operating algorithms may have to decide whether to record or destroy data based on quality of inputs and outputs to avoid large amounts of spurious and insignificant data occupying many TB of storage systems. In summary there are two separate challenges that should not be conflated. There is both a) the technical challenge of delivering online feedback and real time data curation and b) the modelling problem of automating knowledge extraction from the complex HEDP data. Researchers should take care to identify what aspects of their specific scientific problem fit into these two categories, and seek appropriate solutions.

The computerised control of experimental parameters increases the convenience for the experimental operators, allows for automatic parameter scans and reduces the likelihood of experimental errors. Enabling experiment automation requires considerable additional investment and effort in the preparation of experimental apparatus and facilities. However, time and resources spent on this endeavour yield large returns once the experiment is fully operational due to the increase in efficiency and productivity.


\section*{Policy Propositions}

To help the field take advantage of the aforementioned new ways of using data, we make a series of suggestions for how educational and research practices might be beneficially changed.

\subsection*{Education}

With the changes in HEDP data acquisition rates and analysis that have been described in this article, it is important to consider whether researchers who are new to the subject have sufficient training in the topics we have discussed.

There are courses in most universities that attempt to teach aspects of the subject; however they are not typically included in the HEDP curriculum. At the time of the writing, only 3 of the 40 Plasma Physics PhD Projects, Programs and Scholarships listed on \url{www.findaphd.com} contain a data science component.

While the subject is easily within reach for plasma physics graduate students, the unfamiliarity with the Bayesian perspective, and the necessary jargon, render it more difficult than necessary for a self-directed approach. We recommend that, \emph{a minima}, a brief introduction\cite{Mitchell1997,Sivia2006,Brunton2019} be included in one of the Advanced Courses, with the opportunity to take an elective course on Bayesian analysis and Uncertainty Quantification. Ideally, a Bayesian analysis module should be included in future HEDP doctoral programmes, built from the cases presented in this Perspective. In addition to including data science in the general curriculum, existing certificate programs focused on both fundamental ML techniques and multi-disciplinary applications of data science could be leveraged by students and professionals alike. Conferences and workshops like the International Conference on Data Driven Plasma Science (which had its second meeting in 2019) and the 2018 APS Mini-Conference on Machine Learning, Data Science, and Artificial Intelligence in Plasma Research\cite{Wang2020} have flourished in the last few years, and these opportunities should be extended to those studying/training.

For all other graduate students and researchers, a complementary approach should be taken, through targeted workshops and schools. We believe that because the field is developing very quickly, it is preferable to consider teaching these topics as part of the existing HEDP workshops that have been established over the last few years for those new to the area. Topics should include Bayesian and frequentist statistics comparisons, UQ, Markov chain Monte Carlo (MCMC), surrogate building, deep neural nets, and optimization techniques for a range of dimensional spaces. Such workshops and meetings may also provide opportunities to engage the general ML community. The HEDP community could also entice ML researchers to collaborate in our field by presenting or organizing focused sessions at ML conferences, releasing datasets, exploring a ML challenge call focused on an application in our field, and pursuing direct research collaborations. Finally we would note that data science skill sets have broad application both in other research areas, as well as in industry. The proposed training is highly transferable and promises to be valuable to students regardless of specific career goals.

\subsection*{Research Practices}

The changing nature of the field means practices within the field will also have to change. Researchers will need to become familiar with methods needed to run large numbers of simulations, and tools for storing larger amounts of data\cite{Thayer2017}. In particular, the field will have to develop data standards so that data is easily compatible between different facilities. Adopting open data practices\cite{jag_dataset_2020} (like F.A.I.R., findability, accessibility, interoperability, and reusability\cite{Wilkinson2016}) wherever possible is also likely to greatly aid collaboration and comparison of data sets, although this will not be possible for all researchers in this area. The importance of these approaches is already well understood by researchers in other fields, for example high energy physics\cite{Albertsson2018} and astronomy\cite{Borne2009}; we foresee a similar level of data curation will soon be necessary in HEDP.

Many of the methods described in this paper require significant computational resources and good synthetic diagnostics, i.e., good simulations of what the data should look like through the actual pipeline the real data goes through. While most research efforts include a computational component, these are often disjoint from the analysis of experimental data. In future it is advisable for experimental time to also have associated funding for computation and analysis, and the development of synthetic diagnostics/data analysis. Diagnostics and experiments should be designed so that collected data can easily be used in conjunction with other shots (e.g. consistent pixel sizes).  It may even be necessary that the commissioning of instruments comes with a corresponding budget to develop tools to simulate data as seen by the device. Data analysis is a core component of project commissioning and planning in many other areas of physics (e.g. the Euclid telescope\cite{Pasian2012}).

High data rates are permitting the probing of low signal-to-noise phenomena e.g. possible beyond Standard Model physics like axions. If laser based accelerators are to play a role in the future of probing new physics in this way, then statistical analysis must be brought up to the same standards as are practiced in High Energy Physics (HEP). In particular we may wish to adopt stringent statistical significance requirements e.g. 5$\sigma$ for any discovery of beyond Standard Model physics\cite{Lyons2013}. Similarly using blinding methods are likely to become more necessary, where researchers deliberately hide some aspect of the labelling of the data from themselves to prevent subconscious bias or p-hacking\cite{Roodman2003}. Using a blinding protocol does also present dangers and challenges; the complexity and bespoke properties common in many laser-plasma experiments make blinding difficult to implement. However, high-rep rates will make blinding strategies much more viable, and may help for analyses where particular outcomes have high psychological significance and/or there are lots of different potentially viable approaches to doing the analysis.


\section*{Conclusion}

The volume of data on plasma physics at the extremes is rapidly growing and offers the potential for dramatically increased speed of scientific advance. These data are typically multi-modal and describe very non-linear systems, making interpretation challenging for humans, but tractable for AI algorithms. Plasma physics is unlikely to reach the extremely high data rates of HEP in the immediate future, and it is fully the case that AI will not solve all problems in the field. Nonetheless, this new approach in the field offers novel ways of working, and new ways to gain insight - we hope practitioners in the field will be able to find applicability for these methods in their research, see Box 1.

\begin{tcolorbox}

\textbf{BOX 1}

\textbf{Key conclusions:}

\begin{enumerate}
  \item The application of machine learning and modern data science methods to extreme plasma physics and HEDP is rapidly growing and is aiding in producing realistic uncertainties on predictions
  \item Higher repetition rate facilities open up a range of new ways of working; data driven discovery, blinding methods, greater reproducibility, automated data taking
  \item Integrating machine learning based approaches into working practices can greatly save money, time, and human effort
  \item AI based tools are now often more successful at optimising non-linear extreme physics systems and comprehending multi-modal data than humans
\end{enumerate}

\textbf{Key recommendations:}

\begin{enumerate}
  \item Researchers should think carefully about how to best use their data: what methods and diagnostics can they use to take the best data, get sensible uncertainties, and coherently combine with other data sets 
    \item Awards of experimental time and instrument construction should also include greater support for uncertainty quantification, building synthetic diagnostics and data analysis
  \item Plasma physics graduate education and national lab training programmes begin to include basic data science courses
  \item Researchers should try where possible to practice open science best practice; making code and data available publicly, using shared data standards between different facilities
\end{enumerate}

\end{tcolorbox}

In conclusion, modern data science has a lot to offer extreme plasma physics and HEDP science; the community must act now to identify which areas it will make the biggest impact in and make resources and training available to make the most of these novel approaches.


%

\begin{acknowledgments}

This Perspective was the result of a meeting at the Lorentz Center, University of Leiden, 13th-17th January 2020. The Lorentz Centre is funded by the Dutch Research Council (NWO) and the University of Leiden. The meeting also had support from the John Fell Oxford University Press (OUP) Research Fund. The organisers are grateful to Tanja Uitbeijerse (Lorentz Center) for facilitating a fruitful and productive meeting.

P.W.H. acknowledges funding from the Engineering and Physical Sciences Research Council. A portion of this work was performed under the auspices of the US Department of Energy by Lawrence Livermore National Laboratory under Contract DE-AC52-07NA27344. J.A.G and G.J.A were supported by LLNL Laboratory Directed Research and Development project 18-SI-002. The paper has LLNL tracking number LLNL-JRNL-811857. This document was prepared as an account of work sponsored by an agency of the United States government. Neither the United States government nor Lawrence Livermore National Security, LLC, nor any of their employees makes any warranty, expressed or implied, or assumes any legal liability or responsibility for the accuracy, completeness, or usefulness of any information, apparatus, product, or process disclosed, or represents that its use would not infringe privately owned rights. Reference herein to any specific commercial product, process, or service by trade name, trademark, manufacturer, or otherwise does not necessarily constitute or imply its endorsement, recommendation, or favoring by the United States government or Lawrence Livermore National Security, LLC. The views and opinions of authors expressed herein do not necessarily state or reflect those of the United States government or Lawrence Livermore National Security, LLC, and shall not be used for advertising or product endorsement  purposes.

Sandia National Laboratories is a multimission laboratory managed and operated by National Technology \& Engineering Solutions of Sandia, LLC, a wholly owned subsidiary of Honeywell International Inc., for the U.S. Department of Energy’s National Nuclear Security Administration under contract DE-NA0003525. This paper describes objective technical results and analysis. Any subjective views or opinions that might be expressed in the paper do not necessarily represent the views of the U.S. Department of Energy or the United States Government.

\end{acknowledgments}

\section*{Author Contributions}

P.W.H., J.A.G and G.J.A  conceived the work and led the writing of the manuscript. All authors contributed to the manuscript and the ideas discussed at the Lorentz Center Meeting.


\nocite{*}

\bibliography{nature_perspectives.bib}

\end{document}